\documentclass[aps,prl,preprint,superscriptaddress]{revtex4-1}
\usepackage{amsfonts}
\usepackage{amssymb}
\usepackage{amsmath}
\usepackage{graphicx}
\usepackage{caption}

\begin{document}

\title{Monitoring the Birth of an Electronic Wavepacket in a Neutral
Molecule with Attosecond Time-Resolved Photoelectron Spectroscopy}

\author{A. Perveaux}
\affiliation{Laboratoire de Chimie Physique, CNRS,
Universit\'e Paris-Sud, F-91405 Orsay, France}
\affiliation{Institut Charles Gerhardt, CNRS,
Universit\'e Montpellier 2, F-34095 Montpellier, France}

\author{D. Lauvergnat}
\affiliation{Laboratoire de Chimie Physique, CNRS,
Universit\'e Paris-Sud, F-91405 Orsay, France}

\author{F. Gatti}
\affiliation{Institut Charles Gerhardt, CNRS,
Universit\'e Montpellier 2, F-34095 Montpellier, France}

\author{G. J. Hal\'asz}
\affiliation{Department of Information Technology,
University of Debrecen, H-4010 Debrecen, PO Box 12, Hungary}

\author{\'A. Vib\'ok}
\email[]{vibok@phys.unideb.hu}
\affiliation{Department of Theoretical Physics,
University of Debrecen, H-4010 Debrecen, PO Box 5, Hungary}

\author{B. Lasorne}
\email[]{blasorne@univ-montp2.fr}
\affiliation{Institut Charles Gerhardt, CNRS,
Universit\'e Montpellier 2, F-34095 Montpellier, France}

\date{\today}

\begin{abstract}
Numerical simulations are presented to validate the possible use of
cutting-edge attosecond time-resolved photoelectron spectroscopy
to observe in real time the creation of an electronic wavepacket and
subsequent electronic motion in a neutral molecule photoexcited by a
UV pump pulse within a few femtoseconds.
\end{abstract}

\maketitle

\begin{figure}[h]
\includegraphics{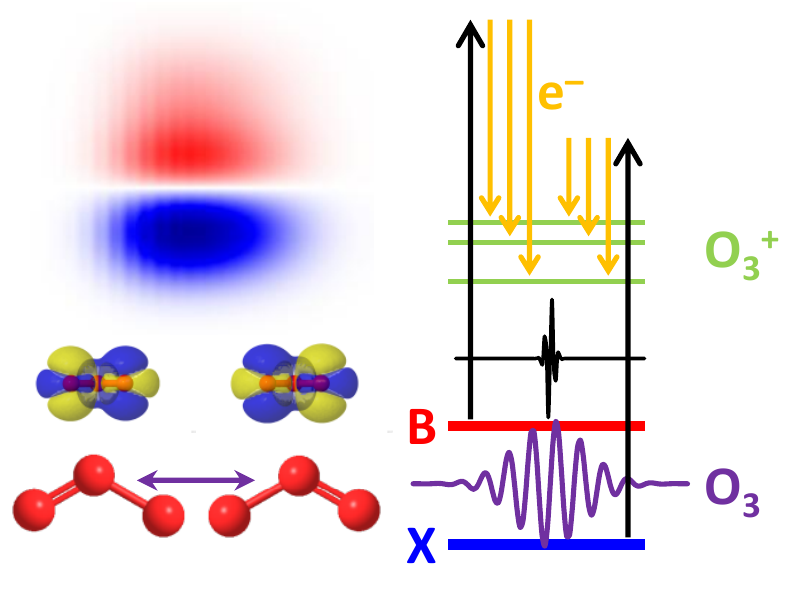}
\caption*{Graphical abstract (for table of content only).}
\end{figure}

\section{Context and Objectives}

The recent emergence of attosecond extreme-ultraviolet (XUV)
pulses \cite{Feri1,Feri2} has given access to observing and controlling
ultrafast electronic motions in real time \cite{Smirnova,Murnane,santra1,
santra2,elefeitosz1,santra3,Vrakking}.
Goulielmakis and coworkers provided the first images of electronic motion in
atoms using an attosecond probe pulse \cite{elefeitosz1}.
Such techniques have thus opened the door to studying molecular processes such
as ultrafast charge migration after sudden ionisation \cite{lenz1a,levine4a}
and ultrafast exciton migration after creating a coherent superposition of
electronic states \cite{mainz2aa,steffi2a,regine1a,Levine5a}.
Experimentally, laser control on the subfemtosecond time scale has already been
evidenced on diatomic molecules: in the dissociative ionisation of the D$_2$
and CO molecules using the carrier envelope phase of a few-cycle
pulse \cite{kling,znakovskaya} to induce charge localisation in the molecular
cation; with the direct manipulation of charge oscillations in
K$_2$ \cite{bayer,wollenhaupt} using selective population of dressed states
within the framework of strong-field temporal control.

Attochemistry, i.e., laser control based on electronic coherence in a neutral
molecule, should soon become a realistic technique, whereby the electronic
motion is steered before the nuclear motion can occur and thus before
intramolecular vibrational relaxation can play a role.
In other words, one should now expect to possess the experimental tools for
observing how chemical bonds really move in a molecular system over the course
of a reaction and further control bond breaking and formation in real time.
Ultrashort sources of coherent light are now approaching the few-femtosecond
regime for UV laser pulses, whereas attosecond pulses are already available in
the XUV domain. One can also expect the emergence of few-cycle UV
subfemtosecond pulses in a near future. However, deciphering experimental
results in this context may prove challenging unless they are supported by an
interpretation based on relevant numerical simulations and adequate
theoretical representations.

Ozone is a well-studied small triatomic molecule which can be handled
theoretically with high precision, and many theoretical and experimental data
such as, e.g., photoelectron spectra or absorption cross sections are available
for it. This molecule, of critical relevance for atmospheric physics and
chemistry, exhibits a high absorption cross section between 200 and 300 nm
followed by dissociation.
This process explains the protective role of the ozone layer
in the Earth atmosphere with respect to UV light from the Sun.
From the experimental point of view, this also implies that ozone can be
efficiently excited electronically by means of few-cycle 3rd harmonic pulses
of Ti:Sa-lasers. In this UV region (called the Hartley band), the absorption
is mostly due to a single electronic excited state (denoted B), such that an
ultrashort UV pulse is expected to create a superposition of the electronic
ground state (X) and the B state almost exclusively, thus making ozone a good
candidate for observing a well-defined electronic wavepacket made of two states.

In previous works, we analysed with accurate quantum dynamics simulations
the coupled electronic and nuclear motion in ozone over 25 fs after
photo-excitation by a 3-fs UV pump pulse \cite{Agnee1a,Agnee2a,Agnee3a}. 
The time evolution of the electronic wavepacket showed a periodic transfer
of electron density between the two bonds, which could be interpreted in
chemical terms as an ultrafast oscillation between both mesomers of ozone
(the two equivalent Lewis structures with localised charges and single or
double bonds on either one or the other side of the molecule).

In the present work, we now concentrate on something different: namely, 
the ultrashort time scales, i.e., the very first few femtoseconds during which
the electronic wavepacket is created. Our objective is to provide some critical
information for the novel experiments that will observe this phenomenon in a
neutral molecule for the first time \cite{Reinhard1}.
In this, the pump pulse should be in the UV domain and no longer than a few
femtoseconds (not to trigger the nuclear motion), while the probe pulse should
be even shorter in order to observe the electronic motion with high-enough
time resolution. The method of choice would thus be time-resolved photoelectron
spectroscopy \cite{spectra1,spectra2,spectra3,Krylov1,Krylov2,serguei1,spanner}
using an attosecond XUV probe pulse.

However, such cutting-edge experiments may not provide clear-cut evidence
for different reasons. Many ionic channels can be accessed upon photoionising
from either X or B, and a number of transitions will overlap within
the same energy window. In addition, the bandwidth of the attosecond probe will
be large: a few eV for a few-hundred-as pulse.
This may prevent any characteristic feature in the spectrum to be observed.
Preliminary numerical simulations are thus essential for future experiments
in order to prove first that contributions from X or B will indeed be
discriminated over time and to identify which energy window is adequate to do
so. To this end, using accurate quantum dynamics and quantum chemistry
calculations, we generated the time-resolved photoelectron spectrum (TRPES)
expected to be observed under such conditions, as further explained below.
Our results clearly show that depletion of X and production of B will be
observed in two distinct energy regions, despite a large bandwidth and
overlapping channels.

\section{Results and Discussion}

\begin{figure}[h]
\includegraphics{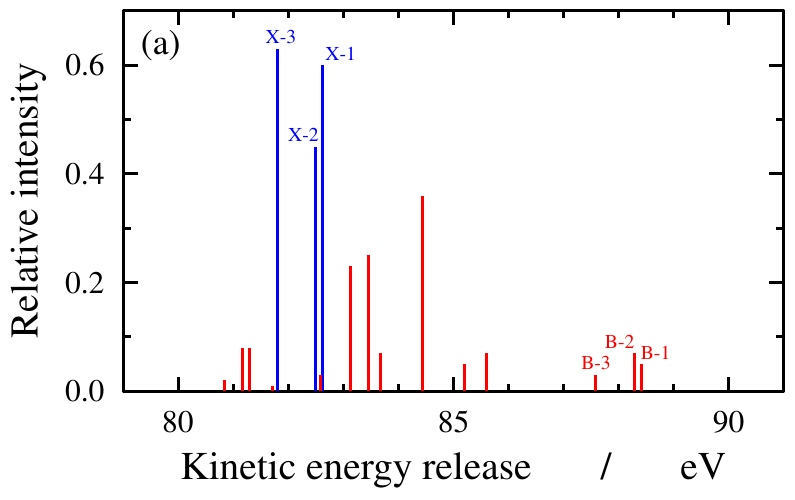}\includegraphics{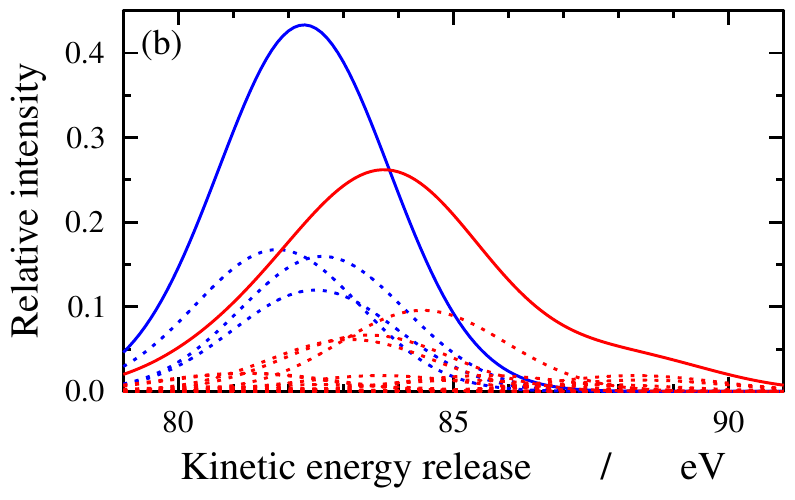} 
\caption{Panel a: Stick photoelectron spectra from X (blue) or B (red) as
functions of the KER for a probe photon energy $E_{ph}=$ 95 eV \cite{Reinhard1};
Panel b: convoluted photoelectron spectra from X (blue) or B (red).
\label{fig:spec}}
\end{figure}

Let us start with the photoelectron spectra obtained by photoionising
from either X or B. First, assuming an ``atomic'' picture (whereby the
rovibrational dynamics and its influence on the structure of the spectrum
are ignored), the corresponding intensities appear as stick spectra
(see Fig. \ref{fig:spec}a), 
\begin{equation}
I_k(\varepsilon)=\sum_j I_{jk}\delta(\varepsilon-\varepsilon_{jk}) ,
\end{equation}
where $\varepsilon$ is the kinetic energy release (KER) of the electron,
with peaks centered at $\epsilon_{jk}=E_k+E_{ph}-E_j$,
$E_{ph}$ being the probe photon energy, and $E_k$ and $E_j$, the energies of
the neutral molecule in state $k=X,B$ and of the cation in state $j$,
respectively (see table \ref{tab:abinitio}).

Assuming one-photon XUV ionisation and using the sudden approximation
\cite{pickup}, the ionisation intensities are near-proportional to the
Dyson norms at the Franck-Condon (FC) point,
$I_{jk}=\left\langle \Phi_{jk}^{D}|\Phi_{jk}^{D}\right\rangle $,
where the Dyson orbitals are defined as  
\begin{equation}
\Phi_{jk}^D(\vec{r})=\sqrt{N}\int d\vec{r}_{1}...d\vec{r}_{N-1}
\psi_k^N(\vec{r}_{1},...\vec{r}_{N}=\vec{r})
\psi_j^{N-1}(\vec{r}_{1},...\vec{r}_{N-1}).
\end{equation}
This approximation is valid when the continuum does not exhibit a rich
structure \cite{spanner}.

The first three states of the cation correspond to intense and narrow
peaks in the photoelectron spectrum when ionising from X \cite{ohtsuka}
(ionisation potentials X-1, X-2, and X-3 around 12-13 eV), thus making them
good candidates for characterising the time evolution of the population of X.
If ionisation also occurs from B, and accounting for an energy shift between
both states of about 6 eV, means that these three peaks may overlap with peaks
corresponding to ionisation from B to cation states around 18-19 eV above X
(between B-12 and B-16; see Table \ref{tab:abinitio}).
We thus limited the KER window to a lower bound of 80.8 eV for a probe
photon at 95.0 eV, thus restricting $j$ to ${1,...,4}$ when ionising
from X and to ${1,...,19}$ from B, up to about 20 eV above X, according to the
energies given in Table \ref{tab:abinitio}. As can be observed in
Fig. \ref{fig:spec}a, both spectra from X and B appear together below 83 eV
(but the peaks from X dominate), whereas only B contributes above 83 eV
with peaks of significant magnitude.

\begin{table}[h]
\caption{\emph{Ab initio} ionisation potentials and Dyson norms at the FC
point with respect to either X or B (electronic structure calculations
performed with the MOLPRO package \cite{molpro-1} at the MRCI/aug-cc-pVQZ
level of theory \cite{MRCI}). The energy difference
between X and B is 5.78 eV. See \cite{ohtsuka} for
a comparison with computed and experimental values.
\label{tab:abinitio}}
\begin{ruledtabular}
\begin{tabular}{|c|c|c|c|c|}
Cation states ($j$)  & $E_j-E_X/eV$  & $E_j-E_B/eV$ & $I_{jX}$ & $I_{jB}$ \\
\hline
 $1(1^{2}A_{1})$  & 12.38  &  6.59 & 0.60 & 0.05 \\
 $2(1^{2}B_{2})$  & 12.51  &  6.72 & 0.45 & 0.07 \\
 $3(1^{2}A_{2})$  & 13.20  &  7.42 & 0.63 & 0.03 \\
 $4(1^{2}B_{1})$  & 14.15  &  8.36 & 0.00 & 0.00 \\
 $5(2^{2}A_{2})$  & 14.45  &  8.66 & 0.00 & 0.00 \\ 
 $6(2^{2}B_{2})$  & 15.18  &  9.40 & 0.01 & 0.07 \\ 
 $7(2^{2}A_{1})$  & 15.58  &  9.80 & 0.00 & 0.05 \\ 
 $8(2^{2}B_{1})$  & 16.35  & 10.56 & 0.19 & 0.36 \\
 $9(3^{2}A_{2})$  & 16.50  & 10.72 & 0.00 & 0.00 \\ 
$10(3^{2}B_{1})$  & 17.10  & 11.32 & 0.08 & 0.07 \\
$11(3^{2}A_{1})$  & 17.32  & 11.54 & 0.20 & 0.25 \\ 
$12(3^{2}B_{2})$  & 17.65  & 11.87 & 0.04 & 0.23 \\ 
$13(4^{2}B_{2})$  & 18.19  & 12.41 & 0.01 & 0.03 \\
$14(4^{2}A_{2})$  & 18.63  & 12.85 & 0.00 & 0.00 \\ 
$15(4^{2}B_{1})$  & 18.61  & 12.83 & 0.00 & 0.00 \\
$16(4^{2}A_{1})$  & 19.07  & 13.29 & 0.00 & 0.01 \\
$17(5^{2}B_{2})$  & 19.61  & 13.83 & 0.03 & 0.08 \\ 
$18(5^{2}A_{1})$  & 19.49  & 13.70 & 0.22 & 0.08 \\ 
$19(6^{2}B_{2})$  & 19.94  & 14.16 & 0.19 & 0.02 \\ 
\end{tabular}
\end{ruledtabular}
\end{table}

Now, such a clear discrimination between both types of spectra could be lost
once accounting for the widths of the peaks. An intrinsic width will be due
to the rovibrational structure for each photoionisation channel, while the
bandwidth of the probe pulse may further enlarge them.
We used a probe pulse of full duration at half-maximum (FDHM) of 500 as,
corresponding to a standard deviation of the intensity $\sigma=$ 1.5 eV in
the energy domain \cite{Reinhard1}.
In this context, the broadening of the spectra due to the bandwidth is
far larger than that due to the rovibrational structure, thus validating the
``atomic'' picture as a first approximation (fixing the geometry at the FC
point for determining the relative intensities of the peaks).
We thus convoluted the stick spectra with a Gaussian window function of
standard deviation $\sigma$, so as to mimic the bandwidth of the XUV
photoionising probe pulse,
\begin{equation}
I_k(\varepsilon)=\frac{1}{\sigma\sqrt{2\pi}}\sum_j
e^{-\frac{(\varepsilon-\varepsilon_{jk})^2}{2\sigma^2}}I_{jk} .
\end{equation}
The resulting energy-resolved spectra are shown on \ref{fig:spec}b.
Despite the strong enlargment, it is to be expected that contributions from
X and B can be discriminated below and above 83 eV, respectively.

Further, to simulate the creation of an electronic wavepacket in the neutral
as a coherent superposition of X and B, we performed quantum dynamics
simulations in the presence of a UV linearly-polarized Gaussian laser pump
pulse with a carrier wavelength, intensity, and FDHM at 260 nm,
10$^{13}$ W/cm$^2$, and 3 fs, respectively \cite{Reinhard1}.
In this, neither the electrons nor the nuclei were in a
stationary state \cite{Agnee1a,Agnee2a,Agnee3a}. 
The multiconfiguration time-dependent Hartree (MCTDH)
method \cite{dieter1a,dieter3a} was applied to solve the time-dependent
nuclear Schr\"odinger equation within a two-electronic-state coupled
representation. The potential energy surfaces and $\vec{R}$-dependent
dipole moments occurring in the radiative coupling terms were taken
from Refs. \cite{schinke1,schinke2}. The FC point
($R_{1}=R_{2}=$ 1.275 $\textrm{\AA}$; $\theta$ = 116.9$^{\circ}$) has
$C_{2v}$ symmetry and, therefore, only the $y$-component ($B_{2}$) of
the transition dipole between X ($^{1}A_{1}$) and B ($^{1}B_{2}$)
is nonzero. Thus, the only effective polarization of the electric
field is $y$.

\begin{figure}[h]
\includegraphics{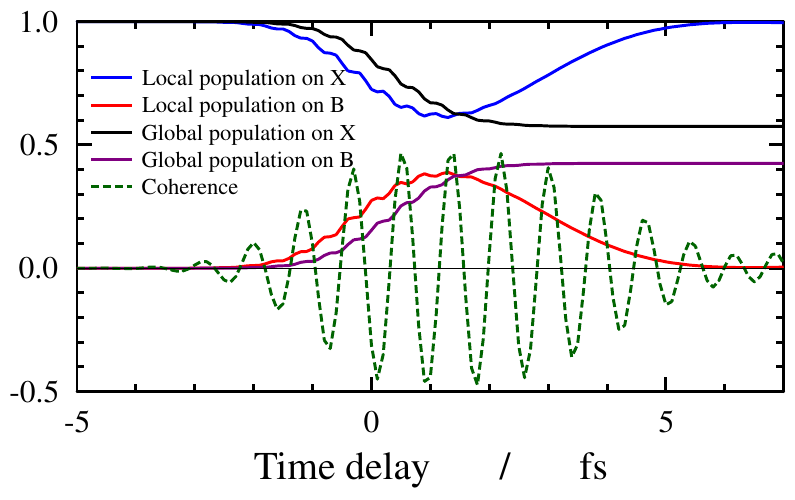} 
\caption{Local X (blue) and B (red) populations and coherence
(green; real part) at the FC point and global X (black)
and B (purple) populations as functions of the time delay
(origin coincident with the maximum of the pump pulse).
\label{fig:pop}}
\end{figure}

Again, using an ``atomic'' picture at the FC point, we define an effective
electronic wavepacket as
\begin{equation}
\left|\psi_{el}(t)\right\rangle =\sum_{k=X,B}c_k(t)
\left|\psi_{el}^{(k)};\vec{R}_{FC}\right\rangle ,
\end{equation}
where $\left|\psi_{el}^{(k)};\vec{R}_{FC}\right\rangle$ are the adiabatic
electronic states of the neutral molecule at the FC point and $c_k(t)$ are the
renormalised nuclear wavepacket components, 
\begin{equation}
c_k(t)=\frac{\psi_{nuc}^{(k)}(\vec{R}_{FC},t)}
{\sqrt{\sum_{l=X,B}\left|\psi_{nuc}^{(l)}(\vec{R}_{FC},t)\right|^2}} .
\end{equation}
The corresponding local populations ($k=k'$) and coherences ($k\neq k'$)
at the FC point are 
$\rho_{kk'}(t)=c_{k}^{*}(t)c_{k'}(t)$, while the global populations are
$P_{k}(t)=\int\ |\psi_{nuc}^{(k)}(\vec{R},t)|^2 d\vec{R}$.
They are plotted on Fig. \ref{fig:pop}, assuming that $t=$ 0 corresponds
to the maximum of the pump pulse (the absolute time, $t$, is thus mapped to the
time delay, $\tau$, between the maxima of the probe and the pump).
As expected, the local populations are similar to the
global ones up to about $\tau=$ 2 fs, which is the focus of the present work.
They further decay while the nuclear wavepackets escape the FC region,
whereas the global populations stay constant.

\begin{figure}
\includegraphics[width=0.5\textwidth]{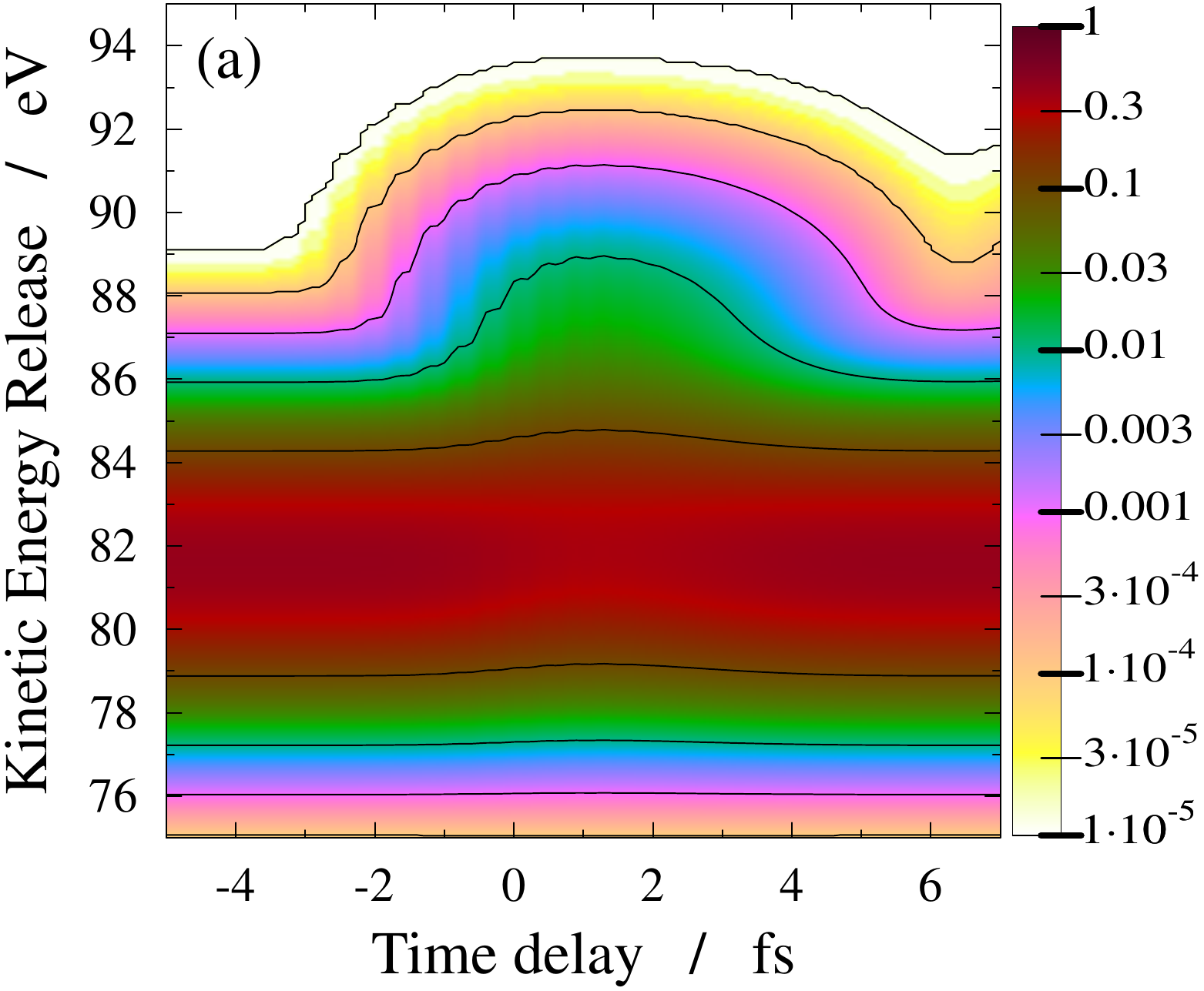}\includegraphics[width=0.5\textwidth]{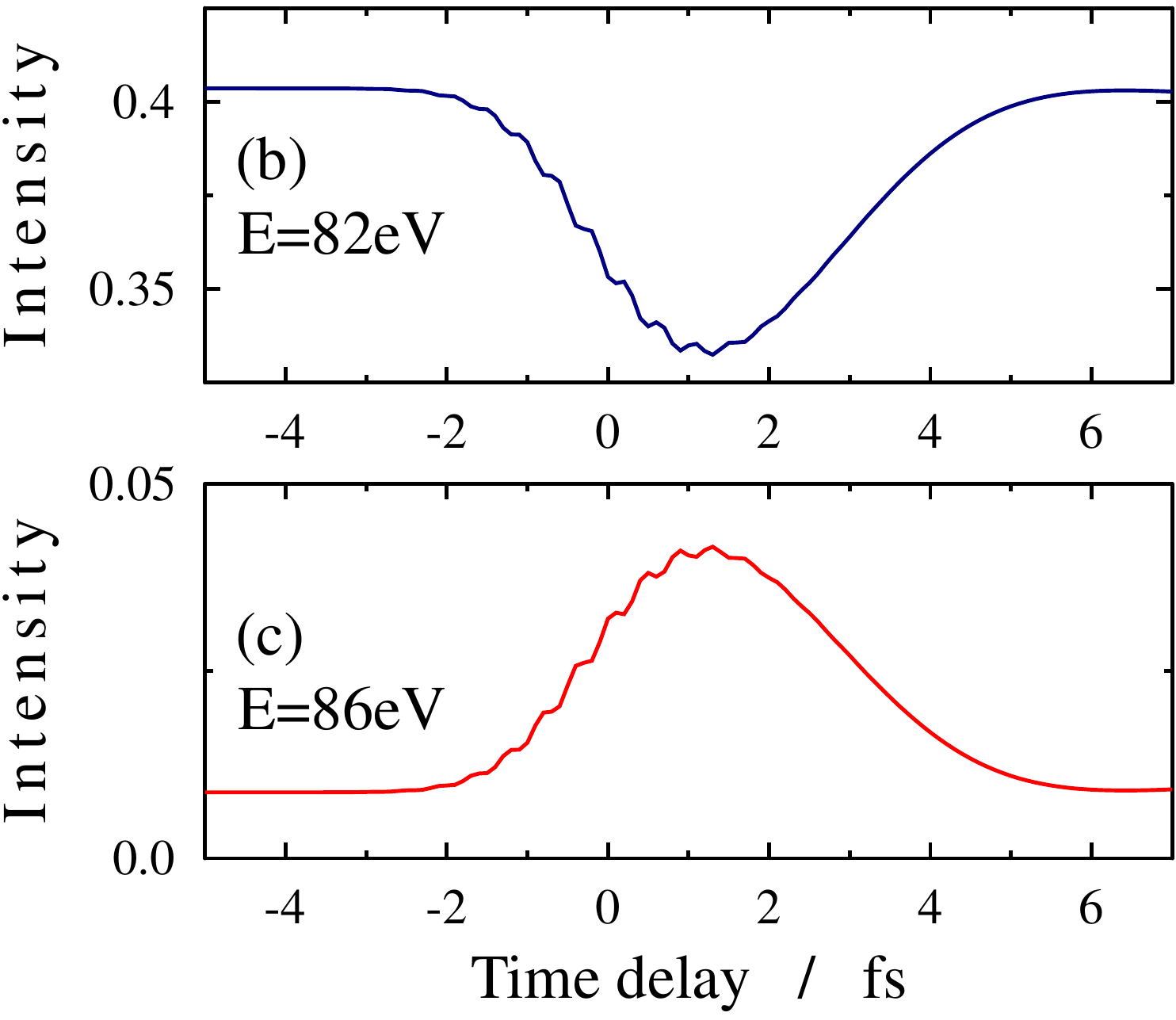}
\caption{Panel a: TRPES (logarithmic scale) as a function
of the time delay (horizontal axis) and kinetic energy release (vertical axis);
panels b and c: cuts against time at fixed energy (linear scale).
\label{fig:trpes}}
\end{figure}

\begin{figure}
\includegraphics[width=0.5\textwidth]{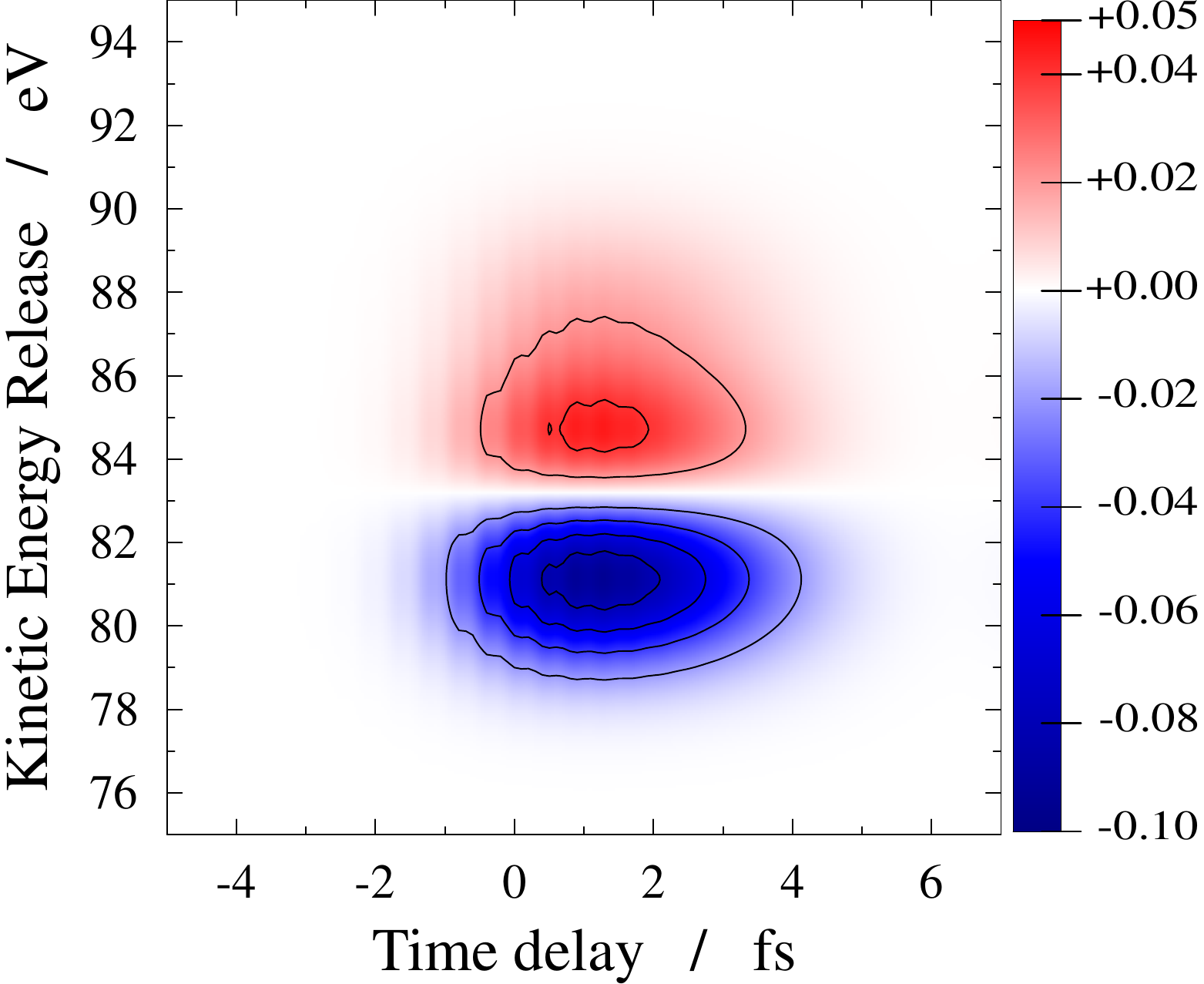}
\caption{Differential TRPES as a function of the time delay
(horizontal axis) and kinetic energy release (vertical axis).
\label{fig:dtrpes}}
\end{figure}

For a stick spectrum, the intensity as a function of the time delay
can be approximated as (for each cation channel):  
\begin{equation}
I_j(\tau)=\sum_{k=X,B}\rho_{kk}(\tau)I_{jk} .
\end{equation}
Using the same convolution procedure as above leads to the following
expression for the TRPES intensity (see Fig. \ref{fig:trpes}),
\begin{equation}
I(\varepsilon,\tau)=\sum_{k=X,B}\rho_{kk}(\tau)I_k(\varepsilon)
=\frac{1}{\sigma\sqrt{2\pi}}\sum_{k=X,B}\rho_{kk}(\tau)
\sum_j e^{-\frac{(\varepsilon-\varepsilon_{jk})^2}{2\sigma^2}}I_{jk} .
\end{equation}
Two significant effects can be noticed on Fig. \ref{fig:trpes}:
\emph{i)} around 82 eV (panel b), between $\tau$ = $-$2 and 2 fs
-- while the pump pulse is on -- the intensity decreases from
0.40 to 0.33 before returning to its original value;
\emph{ii)} around 86 eV (panel c) the intensity increases from
0.01 to 0.04 and returns to 0.01 during the same delay time intervals.
Better contrast is obtained by considering the differential TRPES obtained by
removing the contribution from pure X at all times (see Fig. \ref{fig:dtrpes})
in a similar fashion to differential optical densities in transient absorption
spectroscopy. The depletion of X and the concommitent production of B are
clearly characterised. 

The photoionisation probabilities actually depend on the molecular geometry.
This should be accounted for by integrating the Dyson norms over the nuclear
coordinates when estimating the TRPES for longer times. The limiting case used
here would keep its validity if the Dyson norms had significant magnitudes in the
FC region only and decreased fast when dissociating the molecule. One would thus
observe directly the escape of the wavepacket in the decay of the TRPES signal.
On the other hand, the other limiting case with homogeneous Dyson norms would
yield a constant signal beyond $\tau=$ 2 fs. The experimental signal is likely
to be somewhere in between.

In any case, monitoring the photoelectron spectrum of such a coherent
superposition of X and B as a function of the time delay is expected to yield a
time-dependent pattern that should reflect depletion of X below 83 eV
and production of B above 83 eV on the ultrashort time scale.
Again, results for $\tau=$ $-$2 to 2 fs are expected to be similar to experiments,
whereas the behaviour beyond this time will depend on how the photoionisation
probability depends on the molecular geometry. Our focus here is the ultrashort
time scale in the context of attophysics and the emerging attochemistry field,
whereby sterring directly the electrons of a molecule is expected to provide
control over reaction mechanisms in a near future. 

\section{Summary and Conclusions}

The central objective of this theoretical work was to determine with numerical
simulations if TRPES experiments will be able to monitor the generation of an
electronic wavepacket in the ozone molecule on its real time scale.
Observing the electronic motion in the neutral before any significant nuclear
motion requires using a few-femtosecond UV pulse as a pump. Getting high-enough
time resolution implies using an XUV attosecond pulse as a probe.
The main limitation of such a measurement is the poor energy resolution due to
the large bandwidth of the photoionising pulse. However, we have shown that two
energy regions can be distinguished: one exhibiting depletion of X and one
where production of B is specifically observed.
Such conclusions await further confirmation from experimental investigations.

\begin{acknowledgments}
\section{Acknowledgments}
The authors would like to thank F. Krausz, R. Kienberger and M. Jobst
for support and for fruitful discussions. We acknowledge R. Schinke
for providing the diabatic potential energy surfaces and the transition
dipole moments and H.-D. Meyer for fruitful discussions. \'A. V. also
acknowledges the T\'AMOP-4.2.4.A/ 2-11/1-2012-0001 `National Excellence
Program' and the OTKA (NN103251) project. Financial support by the
CNRS-MTA is greatfully acknowledged.
\end{acknowledgments}

\bibliography{ozone-prl}

\begin{thebibliography}{10}%
\makeatletter
\providecommand \@ifxundefined [1]{%
 \ifx #1\undefined \expandafter \@firstoftwo
 \else \expandafter \@secondoftwo
\fi
}%
\providecommand \@ifnum [1]{%
 \ifnum #1\expandafter \@firstoftwo
 \else \expandafter \@secondoftwo
\fi
}%
\providecommand \enquote [1]{``#1''}%
\providecommand \bibnamefont  [1]{#1}%
\providecommand \bibfnamefont [1]{#1}%
\providecommand \citenamefont [1]{#1}%
\providecommand\href[0]{\@sanitize\@href}%
\providecommand\@href[1]{\endgroup\@@startlink{#1}\endgroup\@@href}%
\providecommand\@@href[1]{#1\@@endlink}%
\providecommand \@sanitize [0]{\begingroup\catcode`\&12\catcode`\#12\relax}%
\@ifxundefined \pdfoutput {\@firstoftwo}{%
 \@ifnum{\z@=\pdfoutput}{\@firstoftwo}{\@secondoftwo}%
}{%
 \providecommand\@@startlink[1]{\leavevmode\special{html:<a href="#1">}}%
 \providecommand\@@endlink[0]{\special{html:</a>}}%
}{%
 \providecommand\@@startlink[1]{%
  \leavevmode
  \pdfstartlink
   attr{/Border[0 0 1 ]/H/I/C[0 1 1]}%
   user{/Subtype/Link/A<</Type/Action/S/URI/URI(#1)>>}%
  \relax
 }%
 \providecommand\@@endlink[0]{\pdfendlink}%
}%
\providecommand \url  [0]{\begingroup\@sanitize \@url }%
\providecommand \@url [1]{\endgroup\@href {#1}{\urlprefix}}%
\providecommand \urlprefix [0]{URL }%
\providecommand \Eprint[0]{\href }%
\@ifxundefined \urlstyle {%
  \providecommand \doi [1]{doi:\discretionary{}{}{}#1}%
}{%
  \providecommand \doi [0]{doi:\discretionary{}{}{}\begingroup
  \urlstyle{rm}\Url }%
}%
\providecommand \doibase [0]{http://dx.doi.org/}%
\providecommand \Doi[1]{\href{\doibase#1}}%
\providecommand \bibAnnote [3]{%
  \BibitemShut{#1}%
  \begin{quotation}\noindent
    \textsc{Key:}\ #2\\\textsc{Annotation:}\ #3%
  \end{quotation}%
}%
\providecommand \bibAnnoteFile [2]{%
  \IfFileExists{#2}{\bibAnnote {#1} {#2} {\input{#2}}}{}%
}%
\providecommand \typeout [0]{\immediate \write \m@ne }%
\providecommand \selectlanguage [0]{\@gobble}%
\providecommand \bibinfo [0]{\@secondoftwo}%
\providecommand \bibfield [0]{\@secondoftwo}%
\providecommand \translation [1]{[#1]}%
\providecommand \BibitemOpen[0]{}%
\providecommand \bibitemStop [0]{}%
\providecommand \bibitemNoStop [0]{.\EOS\space}%
\providecommand \EOS [0]{\spacefactor3000\relax}%
\providecommand \BibitemShut [1]{\csname bibitem#1\endcsname}%
\bibitem{Feri1}%
  \BibitemOpen
  \bibfield{author}{%
  \bibinfo {author} {\bibfnamefont{P.~B.}\ \bibnamefont{Corkum}}\ and\ \bibinfo
  {author} {\bibfnamefont{F.}~\bibnamefont{Krausz}},\ }%
  \bibfield{journal}{%
  \Doi{10.1038/nphys620}{\bibinfo {journal} {Nature Physics}}\ }%
  \textbf{\bibinfo {volume} {3}},\ \bibinfo {pages} {381} (\bibinfo {month}
  {Jun}\ \bibinfo {year} {2007}),\ \url{http://dx.doi.org/10.1038/nphys620}%
  \bibAnnoteFile{NoStop}{Feri1}%
\bibitem{Feri2}%
  \BibitemOpen
  \bibfield{author}{%
  \bibinfo {author} {\bibfnamefont{F.}~\bibnamefont{Krausz}}\ and\ \bibinfo
  {author} {\bibfnamefont{M.}~\bibnamefont{Ivanov}},\ }%
  \bibfield{journal}{%
  \Doi{10.1103/RevModPhys.81.163}{\bibinfo {journal} {Rev. Mod. Phys.}}\ }%
  \textbf{\bibinfo {volume} {81}},\ \bibinfo {pages} {163} (\bibinfo {month}
  {Feb}\ \bibinfo {year} {2009}),\
  \url{http://link.aps.org/doi/10.1103/RevModPhys.81.163}%
  \bibAnnoteFile{NoStop}{Feri2}%
\bibitem{Smirnova}%
  \BibitemOpen
  \bibfield{author}{%
  \bibinfo {author} {\bibfnamefont{O.}~\bibnamefont{Smirnova}}, \bibinfo
  {author} {\bibfnamefont{Y.}~\bibnamefont{Mairesse}}, \bibinfo {author}
  {\bibfnamefont{S.}~\bibnamefont{Patchkovskii}}, \bibinfo {author}
  {\bibfnamefont{N.}~\bibnamefont{Dudovich}}, \bibinfo {author}
  {\bibfnamefont{D.}~\bibnamefont{Villeneuve}}, \bibinfo {author}
  {\bibfnamefont{P.}~\bibnamefont{Corkum}},\ and\ \bibinfo {author}
  {\bibfnamefont{M.~Y.}\ \bibnamefont{Ivanov}},\ }%
  \bibfield{journal}{%
  \Doi{10.1038/nature08253}{\bibinfo {journal} {Nature}}\ }%
  \textbf{\bibinfo {volume} {460}},\ \bibinfo {pages} {972} (\bibinfo {month}
  {Aug}\ \bibinfo {year} {2009}),\ \url{http://dx.doi.org/10.1038/nature08253}%
  \bibAnnoteFile{NoStop}{Smirnova}%
\bibitem{Murnane}%
  \BibitemOpen
  \bibfield{author}{%
  \bibinfo {author} {\bibfnamefont{X.}~\bibnamefont{Zhou}}, \bibinfo {author}
  {\bibfnamefont{P.}~\bibnamefont{Ranitovic}}, \bibinfo {author}
  {\bibfnamefont{C.~W.}\ \bibnamefont{Hogle}}, \bibinfo {author}
  {\bibfnamefont{J.~H.~D.}\ \bibnamefont{Eland}}, \bibinfo {author}
  {\bibfnamefont{H.~C.}\ \bibnamefont{Kapteyn}},\ and\ \bibinfo {author}
  {\bibfnamefont{M.~M.}\ \bibnamefont{Murnane}},\ }%
  \bibfield{journal}{%
  \Doi{10.1038/nphys2211x}{\bibinfo {journal} {Nature Physics}}\ }%
  \textbf{\bibinfo {volume} {8}},\ \bibinfo {pages} {232} (\bibinfo {month}
  {Mar}\ \bibinfo {year} {2012}),\ \url{http://dx.doi.org/10.1038/nphys2211}%
  \bibAnnoteFile{NoStop}{Murnane}%
\bibitem{santra1}%
  \BibitemOpen
  \bibfield{author}{%
  \bibinfo {author} {\bibfnamefont{N.}~\bibnamefont{Rohringer}}, \bibinfo
  {author} {\bibfnamefont{A.}~\bibnamefont{Gordon}},\ and\ \bibinfo {author}
  {\bibfnamefont{R.}~\bibnamefont{Santra}},\ }%
  \bibfield{journal}{%
  \Doi{10.1103/PhysRevA.74.043420}{\bibinfo {journal} {Phys. Rev. A}}\ }%
  \textbf{\bibinfo {volume} {74}},\ \bibinfo {pages} {043420} (\bibinfo {month}
  {Oct}\ \bibinfo {year} {2006}),\
  \url{http://link.aps.org/doi/10.1103/PhysRevA.74.043420}%
  \bibAnnoteFile{NoStop}{santra1}%
\bibitem{santra2}%
  \BibitemOpen
  \bibfield{author}{%
  \bibinfo {author} {\bibfnamefont{N.}~\bibnamefont{Rohringer}}\ and\ \bibinfo
  {author} {\bibfnamefont{R.}~\bibnamefont{Santra}},\ }%
  \bibfield{journal}{%
  \Doi{10.1103/PhysRevA.79.053402}{\bibinfo {journal} {Phys. Rev. A}}\ }%
  \textbf{\bibinfo {volume} {79}},\ \bibinfo {pages} {053402} (\bibinfo {month}
  {May}\ \bibinfo {year} {2009}),\
  \url{http://link.aps.org/doi/10.1103/PhysRevA.79.053402}%
  \bibAnnoteFile{NoStop}{santra2}%
\bibitem{elefeitosz1}%
  \BibitemOpen
  \bibfield{author}{%
  \bibinfo {author} {\bibfnamefont{E.}~\bibnamefont{Goulielmakis}}, \bibinfo
  {author} {\bibfnamefont{Z.-H.}\ \bibnamefont{Loh}}, \bibinfo {author}
  {\bibfnamefont{A.}~\bibnamefont{Wirth}}, \bibinfo {author}
  {\bibfnamefont{R.}~\bibnamefont{Santra}}, \bibinfo {author}
  {\bibfnamefont{N.}~\bibnamefont{Rohringer}}, \bibinfo {author}
  {\bibfnamefont{V.~S.}\ \bibnamefont{Yakovlev}}, \bibinfo {author}
  {\bibfnamefont{S.}~\bibnamefont{Zherebtsov}}, \bibinfo {author}
  {\bibfnamefont{T.}~\bibnamefont{Pfeifer}}, \bibinfo {author}
  {\bibfnamefont{A.~M.}\ \bibnamefont{Azzeer}}, \bibinfo {author}
  {\bibfnamefont{M.~F.}\ \bibnamefont{Kling}}, \bibinfo {author}
  {\bibfnamefont{S.~R.}\ \bibnamefont{Leone}},\ and\ \bibinfo {author}
  {\bibfnamefont{F.}~\bibnamefont{Krausz}},\ }%
  \bibfield{journal}{%
  \Doi{10.1038/nature09212}{\bibinfo {journal} {Nature}}\ }%
  \textbf{\bibinfo {volume} {466}},\ \bibinfo {pages} {739} (\bibinfo {month}
  {Aug}\ \bibinfo {year} {2010}),\ \url{http://dx.doi.org/10.1038/nature09212}%
  \bibAnnoteFile{NoStop}{elefeitosz1}%
\bibitem{santra3}%
  \BibitemOpen
  \bibfield{author}{%
  \bibinfo {author} {\bibfnamefont{G.}~\bibnamefont{Dixit}}, \bibinfo {author}
  {\bibfnamefont{O.}~\bibnamefont{Vendrell}},\ and\ \bibinfo {author}
  {\bibfnamefont{R.}~\bibnamefont{Santra}},\ }%
  \bibfield{journal}{%
  \Doi{10.1073/pnas.1202226109}{\bibinfo {journal} {Proceedings of the National
  Academy of Sciences}}\ }%
  \textbf{\bibinfo {volume} {109}},\ \bibinfo {pages} {11636} (\bibinfo {year}
  {2012}),\ \url{http://www.pnas.org/content/109/29/11636.abstract}%
  \bibAnnoteFile{NoStop}{santra3}%
\bibitem{Vrakking}%
  \BibitemOpen
  \bibfield{author}{%
  \bibinfo {author} {\bibfnamefont{C.}~\bibnamefont{Neidel}}, \bibinfo {author}
  {\bibfnamefont{J.}~\bibnamefont{Klei}}, \bibinfo {author}
  {\bibfnamefont{C.-H.}\ \bibnamefont{Yang}}, \bibinfo {author}
  {\bibfnamefont{A.}~\bibnamefont{Rouz\'ee}}, \bibinfo {author}
  {\bibfnamefont{M.~J.~J.}\ \bibnamefont{Vrakking}}, \bibinfo {author}
  {\bibfnamefont{K.}~\bibnamefont{Kl\"under}}, \bibinfo {author}
  {\bibfnamefont{M.}~\bibnamefont{Miranda}}, \bibinfo {author}
  {\bibfnamefont{C.~L.}\ \bibnamefont{Arnold}}, \bibinfo {author}
  {\bibfnamefont{T.}~\bibnamefont{Fordell}}, \bibinfo {author}
  {\bibfnamefont{A.}~\bibnamefont{L'Huillier}}, \bibinfo {author}
  {\bibfnamefont{M.}~\bibnamefont{Gisselbrecht}}, \bibinfo {author}
  {\bibfnamefont{P.}~\bibnamefont{Johnsson}}, \bibinfo {author}
  {\bibfnamefont{M.~P.}\ \bibnamefont{Dinh}}, \bibinfo {author}
  {\bibfnamefont{E.}~\bibnamefont{Suraud}}, \bibinfo {author}
  {\bibfnamefont{P.-G.}\ \bibnamefont{Reinhard}}, \bibinfo {author}
  {\bibfnamefont{V.}~\bibnamefont{Despr\'e}}, \bibinfo {author}
  {\bibfnamefont{M.~A.~L.}\ \bibnamefont{Marques}},\ and\ \bibinfo {author}
  {\bibfnamefont{F.}~\bibnamefont{L\'epine}},\ }%
  \bibfield{journal}{%
  \Doi{10.1103/PhysRevLett.111.033001}{\bibinfo {journal} {Phys. Rev. Lett.}}\
  }%
  \textbf{\bibinfo {volume} {111}},\ \bibinfo {pages} {033001} (\bibinfo
  {month} {Jul}\ \bibinfo {year} {2013}),\
  \url{http://link.aps.org/doi/10.1103/PhysRevLett.111.033001}%
  \bibAnnoteFile{NoStop}{Vrakking}%
\bibitem{lenz1a}%
  \BibitemOpen
  \bibfield{author}{%
  \bibinfo {author} {\bibfnamefont{A.~I.}\ \bibnamefont{Kuleff}}, \bibinfo
  {author} {\bibfnamefont{J.}~\bibnamefont{Breidbach}},\ and\ \bibinfo {author}
  {\bibfnamefont{L.~S.}\ \bibnamefont{Cederbaum}},\ }%
  \bibfield{journal}{%
  \Doi{10.1063/1.1961341}{\bibinfo {journal} {The Journal of Chemical
  Physics}}\ }%
  \textbf{\bibinfo {volume} {123}},\ \bibinfo {eid} {044111} (\bibinfo {year}
  {2005}),\
  \url{http://scitation.aip.org/content/aip/journal/jcp/123/4/10.1063/1.196134%
1}%
  \bibAnnoteFile{NoStop}{lenz1a}%
\bibitem{levine4a}%
  \BibitemOpen
  \bibfield{author}{%
  \bibinfo {author} {\bibfnamefont{F.}~\bibnamefont{Remacle}}\ and\ \bibinfo
  {author} {\bibfnamefont{R.~D.}\ \bibnamefont{Levine}},\ }%
  \bibfield{journal}{%
  \Doi{10.1103/PhysRevA.83.013411}{\bibinfo {journal} {Phys. Rev. A}}\ }%
  \textbf{\bibinfo {volume} {83}},\ \bibinfo {pages} {013411} (\bibinfo {month}
  {Jan}\ \bibinfo {year} {2011}),\
  \url{http://link.aps.org/doi/10.1103/PhysRevA.83.013411}%
  \bibAnnoteFile{NoStop}{levine4a}%
\bibitem{mainz2aa}%
  \BibitemOpen
  \bibfield{author}{%
  \bibinfo {author} {\bibfnamefont{A.~D.}\ \bibnamefont{Bandrauk}}, \bibinfo
  {author} {\bibfnamefont{S.}~\bibnamefont{Chelkowski}}, \bibinfo {author}
  {\bibfnamefont{P.~B.}\ \bibnamefont{Corkum}}, \bibinfo {author}
  {\bibfnamefont{J.}~\bibnamefont{Manz}},\ and\ \bibinfo {author}
  {\bibfnamefont{G.~L.}\ \bibnamefont{Yudin}},\ }%
  \bibfield{journal}{%
  \Doi{10.1088/0953-4075/42/13/134001}{\bibinfo {journal} {Journal of Physics
  B: Atomic, Molecular and Optical Physics}}\ }%
  \textbf{\bibinfo {volume} {42}},\ \bibinfo {pages} {134001} (\bibinfo {year}
  {2009}),\ \url{http://stacks.iop.org/0953-4075/42/i=13/a=134001}%
  \bibAnnoteFile{NoStop}{mainz2aa}%
\bibitem{steffi2a}%
  \BibitemOpen
  \bibfield{author}{%
  \bibinfo {author} {\bibfnamefont{S.}~\bibnamefont{Gr\"afe}}, \bibinfo
  {author} {\bibfnamefont{V.}~\bibnamefont{Engel}},\ and\ \bibinfo {author}
  {\bibfnamefont{M.~Y.}\ \bibnamefont{Ivanov}},\ }%
  \bibfield{journal}{%
  \Doi{10.1103/PhysRevLett.101.103001}{\bibinfo {journal} {Phys. Rev. Lett.}}\
  }%
  \textbf{\bibinfo {volume} {101}},\ \bibinfo {pages} {103001} (\bibinfo
  {month} {Sep}\ \bibinfo {year} {2008}),\
  \url{http://link.aps.org/doi/10.1103/PhysRevLett.101.103001}%
  \bibAnnoteFile{NoStop}{steffi2a}%
\bibitem{regine1a}%
  \BibitemOpen
  \bibfield{author}{%
  \bibinfo {author} {\bibfnamefont{D.}~\bibnamefont{Geppert}}, \bibinfo
  {author} {\bibfnamefont{P.}~\bibnamefont{von~den Hoff}},\ and\ \bibinfo
  {author} {\bibfnamefont{R.}~\bibnamefont{de~Vivie-Riedle}},\ }%
  \bibfield{journal}{%
  \Doi{10.1088/0953-4075/41/7/074006}{\bibinfo {journal} {Journal of Physics B:
  Atomic, Molecular and Optical Physics}}\ }%
  \textbf{\bibinfo {volume} {41}},\ \bibinfo {pages} {074006} (\bibinfo {year}
  {2008}),\ \url{http://stacks.iop.org/0953-4075/41/i=7/a=074006}%
  \bibAnnoteFile{NoStop}{regine1a}%
\bibitem{Levine5a}%
  \BibitemOpen
  \bibfield{author}{%
  \bibinfo {author} {\bibfnamefont{B.}~\bibnamefont{Mignolet}}, \bibinfo
  {author} {\bibfnamefont{R.~D.}\ \bibnamefont{Levine}},\ and\ \bibinfo
  {author} {\bibfnamefont{F.}~\bibnamefont{Remacle}},\ }%
  \bibfield{journal}{%
  \Doi{10.1103/PhysRevA.86.053429}{\bibinfo {journal} {Phys. Rev. A}}\ }%
  \textbf{\bibinfo {volume} {86}},\ \bibinfo {pages} {053429} (\bibinfo {month}
  {Nov}\ \bibinfo {year} {2012}),\
  \url{http://link.aps.org/doi/10.1103/PhysRevA.86.053429}%
  \bibAnnoteFile{NoStop}{Levine5a}%
\bibitem{kling}%
  \BibitemOpen
  \bibfield{author}{%
  \bibinfo {author} {\bibfnamefont{M.~F.}\ \bibnamefont{Kling}}, \bibinfo
  {author} {\bibfnamefont{C.}~\bibnamefont{Siedschlag}}, \bibinfo {author}
  {\bibfnamefont{A.~J.}\ \bibnamefont{Verhoef}}, \bibinfo {author}
  {\bibfnamefont{J.~I.}\ \bibnamefont{Khan}}, \bibinfo {author}
  {\bibfnamefont{M.}~\bibnamefont{Schultze}}, \bibinfo {author}
  {\bibfnamefont{T.}~\bibnamefont{Uphues}}, \bibinfo {author}
  {\bibfnamefont{Y.}~\bibnamefont{Ni}}, \bibinfo {author}
  {\bibfnamefont{M.}~\bibnamefont{Uiberacker}}, \bibinfo {author}
  {\bibfnamefont{M.}~\bibnamefont{Drescher}}, \bibinfo {author}
  {\bibfnamefont{F.}~\bibnamefont{Krausz}},\ and\ \bibinfo {author}
  {\bibfnamefont{M.~J.~J.}\ \bibnamefont{Vrakking}},\ }%
  \bibfield{journal}{%
  \Doi{10.1126/science.1126259}{\bibinfo {journal} {Science}}\ }%
  \textbf{\bibinfo {volume} {312}},\ \bibinfo {pages} {246} (\bibinfo {year}
  {2006}),\ \url{http://www.sciencemag.org/content/312/5771/246.abstract}%
  \bibAnnoteFile{NoStop}{kling}%
\bibitem{znakovskaya}%
  \BibitemOpen
  \bibfield{author}{%
  \bibinfo {author} {\bibfnamefont{I.}~\bibnamefont{Znakovskaya}}, \bibinfo
  {author} {\bibfnamefont{P.}~\bibnamefont{von~den Hoff}}, \bibinfo {author}
  {\bibfnamefont{S.}~\bibnamefont{Zherebtsov}}, \bibinfo {author}
  {\bibfnamefont{A.}~\bibnamefont{Wirth}}, \bibinfo {author}
  {\bibfnamefont{O.}~\bibnamefont{Herrwerth}}, \bibinfo {author}
  {\bibfnamefont{M.~J.~J.}\ \bibnamefont{Vrakking}}, \bibinfo {author}
  {\bibfnamefont{R.}~\bibnamefont{de~Vivie-Riedle}},\ and\ \bibinfo {author}
  {\bibfnamefont{M.~F.}\ \bibnamefont{Kling}},\ }%
  \bibfield{journal}{%
  \Doi{10.1103/PhysRevLett.103.103002}{\bibinfo {journal} {Phys. Rev. Lett.}}\
  }%
  \textbf{\bibinfo {volume} {103}},\ \bibinfo {pages} {103002} (\bibinfo
  {month} {Sep}\ \bibinfo {year} {2009}),\
  \url{http://link.aps.org/doi/10.1103/PhysRevLett.103.103002}%
  \bibAnnoteFile{NoStop}{znakovskaya}%
\bibitem{bayer}%
  \BibitemOpen
  \bibfield{author}{%
  \bibinfo {author} {\bibfnamefont{T.}~\bibnamefont{Bayer}}, \bibinfo {author}
  {\bibfnamefont{H.}~\bibnamefont{Braun}}, \bibinfo {author}
  {\bibfnamefont{C.}~\bibnamefont{Sarpe}}, \bibinfo {author}
  {\bibfnamefont{R.}~\bibnamefont{Siemering}}, \bibinfo {author}
  {\bibfnamefont{P.}~\bibnamefont{von~den Hoff}}, \bibinfo {author}
  {\bibfnamefont{R.}~\bibnamefont{de~Vivie-Riedle}}, \bibinfo {author}
  {\bibfnamefont{T.}~\bibnamefont{Baumert}},\ and\ \bibinfo {author}
  {\bibfnamefont{M.}~\bibnamefont{Wollenhaupt}},\ }%
  \bibfield{journal}{%
  \Doi{10.1103/PhysRevLett.110.123003}{\bibinfo {journal} {Phys. Rev. Lett.}}\
  }%
  \textbf{\bibinfo {volume} {110}},\ \bibinfo {pages} {123003} (\bibinfo
  {month} {Mar}\ \bibinfo {year} {2013}),\
  \url{http://link.aps.org/doi/10.1103/PhysRevLett.110.123003}%
  \bibAnnoteFile{NoStop}{bayer}%
\bibitem{wollenhaupt}%
  \BibitemOpen
  \bibfield{author}{%
  \bibinfo {author} {\bibfnamefont{M.}~\bibnamefont{Wollenhaupt}}\ and\
  \bibinfo {author} {\bibfnamefont{T.}~\bibnamefont{Baumert}},\ }%
  \bibfield{journal}{%
  \Doi{10.1016/j.jphotochem.2006.03.010}{\bibinfo {journal} {Journal of
  Photochemistry and Photobiology A: Chemistry}}\ }%
  \textbf{\bibinfo {volume} {180}},\ \bibinfo {pages} {248} (\bibinfo {year}
  {2006}),\
  \url{http://www.sciencedirect.com/science/article/pii/S1010603006001444}%
  \bibAnnoteFile{NoStop}{wollenhaupt}%
\bibitem{Agnee1a}%
  \BibitemOpen
  \bibfield{author}{%
  \bibinfo {author} {\bibfnamefont{G.~J.}\ \bibnamefont{Hal\'asz}}, \bibinfo
  {author} {\bibfnamefont{A.}~\bibnamefont{Perveaux}}, \bibinfo {author}
  {\bibfnamefont{B.}~\bibnamefont{Lasorne}}, \bibinfo {author}
  {\bibfnamefont{M.~A.}\ \bibnamefont{Robb}}, \bibinfo {author}
  {\bibfnamefont{F.}~\bibnamefont{Gatti}},\ and\ \bibinfo {author}
  {\bibfnamefont{A.}~\bibnamefont{Vib\'ok}},\ }%
  \bibfield{journal}{%
  \Doi{10.1103/PhysRevA.86.043426}{\bibinfo {journal} {Phys. Rev. A}}\ }%
  \textbf{\bibinfo {volume} {86}},\ \bibinfo {pages} {043426} (\bibinfo {month}
  {Oct}\ \bibinfo {year} {2012}),\
  \url{http://link.aps.org/doi/10.1103/PhysRevA.86.043426}%
  \bibAnnoteFile{NoStop}{Agnee1a}%
\bibitem{Agnee2a}%
  \BibitemOpen
  \bibfield{author}{%
  \bibinfo {author} {\bibfnamefont{G.~J.}\ \bibnamefont{Hal\'asz}}, \bibinfo
  {author} {\bibfnamefont{A.}~\bibnamefont{Perveaux}}, \bibinfo {author}
  {\bibfnamefont{B.}~\bibnamefont{Lasorne}}, \bibinfo {author}
  {\bibfnamefont{M.~A.}\ \bibnamefont{Robb}}, \bibinfo {author}
  {\bibfnamefont{F.}~\bibnamefont{Gatti}},\ and\ \bibinfo {author}
  {\bibfnamefont{A.}~\bibnamefont{Vib\'ok}},\ }%
  \bibfield{journal}{%
  \Doi{10.1103/PhysRevA.88.023425}{\bibinfo {journal} {Phys. Rev. A}}\ }%
  \textbf{\bibinfo {volume} {88}},\ \bibinfo {pages} {023425} (\bibinfo {month}
  {Aug}\ \bibinfo {year} {2013}),\
  \url{http://link.aps.org/doi/10.1103/PhysRevA.88.023425}%
  \bibAnnoteFile{NoStop}{Agnee2a}%
\bibitem{Agnee3a}%
  \BibitemOpen
  \bibfield{author}{%
  \bibinfo {author} {\bibfnamefont{A.}~\bibnamefont{Perveaux}}, \bibinfo
  {author} {\bibfnamefont{D.}~\bibnamefont{Lauvergnat}}, \bibinfo {author}
  {\bibfnamefont{B.}~\bibnamefont{Lasorne}}, \bibinfo {author}
  {\bibfnamefont{F.}~\bibnamefont{Gatti}}, \bibinfo {author}
  {\bibfnamefont{M.~A.}\ \bibnamefont{Robb}}, \bibinfo {author}
  {\bibfnamefont{G.~J.}\ \bibnamefont{Hal\'asz}},\ and\ \bibinfo {author}
  {\bibfnamefont{A.}~\bibnamefont{Vib\'ok}},\ }%
  \bibfield{journal}{%
  \Doi{10.1088/0953-4075/47/12/124010}{\bibinfo {journal} {Journal of Physics
  B: Atomic, Molecular and Optical Physics}}\ }%
  \textbf{\bibinfo {volume} {47}},\ \bibinfo {pages} {124010} (\bibinfo {year}
  {2014}),\ \url{http://stacks.iop.org/0953-4075/47/i=12/a=124010}%
  \bibAnnoteFile{NoStop}{Agnee3a}%
\bibitem{Reinhard1}%
  \BibitemOpen
  \bibfield{author}{%
  \bibinfo {author} {\bibfnamefont{R.}~\bibnamefont{Kienberger}}, \bibinfo
  {author} {\bibfnamefont{M.}~\bibnamefont{Jobst}},\ and\ \bibinfo {author}
  {\bibfnamefont{F.}~\bibnamefont{Krausz}},\ }%
  \bibinfo {howpublished} {private communication}%
  \bibAnnoteFile{NoStop}{Reinhard1}%
\bibitem{spectra1}%
  \BibitemOpen
  \bibfield{author}{%
  \bibinfo {author} {\bibfnamefont{M.}~\bibnamefont{Wollenhaupt}}, \bibinfo
  {author} {\bibfnamefont{V.}~\bibnamefont{Engel}},\ and\ \bibinfo {author}
  {\bibfnamefont{T.}~\bibnamefont{Baumert}},\ }%
  \bibfield{journal}{%
  \Doi{10.1146/annurev.physchem.56.092503.141315}{\bibinfo {journal} {Annual
  Review of Physical Chemistry}}\ }%
  \textbf{\bibinfo {volume} {56}},\ \bibinfo {pages} {25} (\bibinfo {year}
  {2005}),\ \url{http://dx.doi.org/10.1146/annurev.physchem.56.092503.141315}%
  \bibAnnoteFile{NoStop}{spectra1}%
\bibitem{spectra2}%
  \BibitemOpen
  \bibfield{author}{%
  \bibinfo {author} {\bibfnamefont{A.}~\bibnamefont{Stolow}},\ }%
  \bibfield{journal}{%
  \Doi{10.1146/annurev.physchem.54.011002.103809}{\bibinfo {journal} {Annual
  Review of Physical Chemistry}}\ }%
  \textbf{\bibinfo {volume} {54}},\ \bibinfo {pages} {89} (\bibinfo {year}
  {2003}),\ \url{http://dx.doi.org/10.1146/annurev.physchem.54.011002.103809}%
  \bibAnnoteFile{NoStop}{spectra2}%
\bibitem{spectra3}%
  \BibitemOpen
  \bibfield{author}{%
  \bibinfo {author} {\bibfnamefont{A.}~\bibnamefont{Stolow}}, \bibinfo {author}
  {\bibfnamefont{A.~E.}\ \bibnamefont{Bragg}},\ and\ \bibinfo {author}
  {\bibfnamefont{D.~M.}\ \bibnamefont{Neumark}},\ }%
  \bibfield{journal}{%
  \Doi{10.1021/cr020683w}{\bibinfo {journal} {Chemical Reviews}}\ }%
  \textbf{\bibinfo {volume} {104}},\ \bibinfo {pages} {1719} (\bibinfo {year}
  {2004}),\ \url{http://pubs.acs.org/doi/abs/10.1021/cr020683w}%
  \bibAnnoteFile{NoStop}{spectra3}%
\bibitem{Krylov1}%
  \BibitemOpen
  \bibfield{author}{%
  \bibinfo {author} {\bibfnamefont{C.}~\bibnamefont{Melania~Oana}}\ and\
  \bibinfo {author} {\bibfnamefont{A.~I.}\ \bibnamefont{Krylov}},\ }%
  \bibfield{journal}{%
  \Doi{10.1063/1.2805393}{\bibinfo {journal} {The Journal of Chemical
  Physics}}\ }%
  \textbf{\bibinfo {volume} {127}},\ \bibinfo {eid} {234106} (\bibinfo {year}
  {2007}),\
  \url{http://scitation.aip.org/content/aip/journal/jcp/127/23/10.1063/1.28053%
93}%
  \bibAnnoteFile{NoStop}{Krylov1}%
\bibitem{Krylov2}%
  \BibitemOpen
  \bibfield{author}{%
  \bibinfo {author} {\bibfnamefont{C.~M.}\ \bibnamefont{Oana}}\ and\ \bibinfo
  {author} {\bibfnamefont{A.~I.}\ \bibnamefont{Krylov}},\ }%
  \bibfield{journal}{%
  \Doi{10.1063/1.3231143}{\bibinfo {journal} {The Journal of Chemical
  Physics}}\ }%
  \textbf{\bibinfo {volume} {131}},\ \bibinfo {eid} {124114} (\bibinfo {year}
  {2009}),\
  \url{http://scitation.aip.org/content/aip/journal/jcp/131/12/10.1063/1.32311%
43}%
  \bibAnnoteFile{NoStop}{Krylov2}%
\bibitem{serguei1}%
  \BibitemOpen
  \bibfield{author}{%
  \bibinfo {author} {\bibfnamefont{M.}~\bibnamefont{Spanner}}\ and\ \bibinfo
  {author} {\bibfnamefont{S.}~\bibnamefont{Patchkovskii}},\ }%
  \bibfield{journal}{%
  \Doi{10.1103/PhysRevA.80.063411}{\bibinfo {journal} {Phys. Rev. A}}\ }%
  \textbf{\bibinfo {volume} {80}},\ \bibinfo {pages} {063411} (\bibinfo {month}
  {Dec}\ \bibinfo {year} {2009}),\
  \url{http://link.aps.org/doi/10.1103/PhysRevA.80.063411}%
  \bibAnnoteFile{NoStop}{serguei1}%
\bibitem{spanner}%
  \BibitemOpen
  \bibfield{author}{%
  \bibinfo {author} {\bibfnamefont{M.}~\bibnamefont{Spanner}}, \bibinfo
  {author} {\bibfnamefont{S.}~\bibnamefont{Patchkovskii}}, \bibinfo {author}
  {\bibfnamefont{C.}~\bibnamefont{Zhou}}, \bibinfo {author}
  {\bibfnamefont{S.}~\bibnamefont{Matsika}}, \bibinfo {author}
  {\bibfnamefont{M.}~\bibnamefont{Kotur}},\ and\ \bibinfo {author}
  {\bibfnamefont{T.~C.}\ \bibnamefont{Weinacht}},\ }%
  \bibfield{journal}{%
  \Doi{10.1103/PhysRevA.86.053406}{\bibinfo {journal} {Phys. Rev. A}}\ }%
  \textbf{\bibinfo {volume} {86}},\ \bibinfo {pages} {053406} (\bibinfo {month}
  {Nov}\ \bibinfo {year} {2012}),\
  \url{http://link.aps.org/doi/10.1103/PhysRevA.86.053406}%
  \bibAnnoteFile{NoStop}{spanner}%
\bibitem{pickup}%
  \BibitemOpen
  \bibfield{author}{%
  \bibinfo {author} {\bibfnamefont{B.~T.}\ \bibnamefont{Pickup}},\ }%
  \bibfield{journal}{%
  \Doi{10.1016/0301-0104(77)85131-8}{\bibinfo {journal} {Chemical Physics}}\ }%
  \textbf{\bibinfo {volume} {19}},\ \bibinfo {pages} {193} (\bibinfo {year}
  {1977}),\
  \url{http://www.sciencedirect.com/science/article/pii/0301010477851318}%
  \bibAnnoteFile{NoStop}{pickup}%
\bibitem{ohtsuka}%
  \BibitemOpen
  \bibfield{author}{%
  \bibinfo {author} {\bibfnamefont{Y.}~\bibnamefont{Ohtsuka}}, \bibinfo
  {author} {\bibfnamefont{J.-y.}\ \bibnamefont{Hasegawa}},\ and\ \bibinfo
  {author} {\bibfnamefont{H.}~\bibnamefont{Nakatsuji}},\ }%
  \bibfield{journal}{%
  \Doi{10.1016/j.chemphys.2006.12.008}{\bibinfo {journal} {Chemical Physics}}\
  }%
  \textbf{\bibinfo {volume} {332}},\ \bibinfo {pages} {262} (\bibinfo {year}
  {2007}),\
  \url{http://www.sciencedirect.com/science/article/pii/S0301010406006537}%
  \bibAnnoteFile{NoStop}{ohtsuka}%
\bibitem{molpro-1}%
  \BibitemOpen
  \bibfield{author}{%
  \bibinfo {author} {\bibfnamefont{H.-J.}\ \bibnamefont{Werner}}, \bibinfo
  {author} {\bibfnamefont{P.~J.}\ \bibnamefont{Knowles}}, \bibinfo {author}
  {\bibfnamefont{G.}~\bibnamefont{Knizia}}, \bibinfo {author}
  {\bibfnamefont{F.~R.}\ \bibnamefont{Manby}}, \bibinfo {author}
  {\bibfnamefont{M.}~\bibnamefont{Sch\"utz}}, \emph{et~al.},\ }%
  \enquote{\bibinfo {title} {Molpro, version 2010.1, a package of ab initio
  programs},}\  (\bibinfo {year} {2010}),\ \bibinfo {note} {see
  http://www.molpro.net}%
  \bibAnnoteFile{NoStop}{molpro-1}%
\bibitem{MRCI}%
  \BibitemOpen
  \bibfield{author}{%
  \bibinfo {author} {\bibfnamefont{P.~J.}\ \bibnamefont{Knowles}}\ and\
  \bibinfo {author} {\bibfnamefont{H.-J.}\ \bibnamefont{Werner}},\ }%
  \bibfield{journal}{%
  \Doi{10.1007/BF01117405}{\bibinfo {journal} {Theoretica Chimica Acta}}\ }%
  \textbf{\bibinfo {volume} {84}},\ \bibinfo {pages} {95} (\bibinfo {year}
  {1992}),\ \url{http://dx.doi.org/10.1007/BF01117405}%
  \bibAnnoteFile{NoStop}{MRCI}%
\bibitem{dieter1a}%
  \BibitemOpen
  \bibfield{author}{%
  \bibinfo {author} {\bibfnamefont{H.-D.}\ \bibnamefont{Meyer}}, \bibinfo
  {author} {\bibfnamefont{U.}~\bibnamefont{Manthe}},\ and\ \bibinfo {author}
  {\bibfnamefont{L.~S.}\ \bibnamefont{Cederbaum}},\ }%
  \bibfield{journal}{%
  \Doi{10.1016/0009-2614(90)87014-I}{\bibinfo {journal} {Chemical Physics
  Letters}}\ }%
  \textbf{\bibinfo {volume} {165}},\ \bibinfo {pages} {73} (\bibinfo {year}
  {1990}),\
  \url{http://www.sciencedirect.com/science/article/pii/000926149087014I}%
  \bibAnnoteFile{NoStop}{dieter1a}%
\bibitem{dieter3a}%
  \BibitemOpen
  \bibfield{author}{%
  \bibinfo {author} {\bibfnamefont{G.~A.}\ \bibnamefont{Worth}}, \bibinfo
  {author} {\bibfnamefont{M.~H.}\ \bibnamefont{Beck}}, \bibinfo {author}
  {\bibfnamefont{A.}~\bibnamefont{J\"ackle}},\ and\ \bibinfo {author}
  {\bibfnamefont{H.-D.}\ \bibnamefont{Meyer}},\ }%
  \bibinfo {howpublished} {The {MCTDH} {P}ackage, {V}ersion 8.2, (2000). H.-D.
  Meyer, {V}ersion 8.3 (2002), {V}ersion 8.4 (2007). {S}ee
  http://mctdh.uni-hd.de}%
  \bibAnnoteFile{NoStop}{dieter3a}%
\bibitem{schinke1}%
  \BibitemOpen
  \bibfield{author}{%
  \bibinfo {author} {\bibfnamefont{Z.-W.}\ \bibnamefont{Qu}}, \bibinfo {author}
  {\bibfnamefont{H.}~\bibnamefont{Zhu}}, \bibinfo {author}
  {\bibfnamefont{S.~Y.}\ \bibnamefont{Grebenshchikov}},\ and\ \bibinfo {author}
  {\bibfnamefont{R.}~\bibnamefont{Schinke}},\ }%
  \bibfield{journal}{%
  \Doi{10.1063/1.2001650}{\bibinfo {journal} {The Journal of Chemical
  Physics}}\ }%
  \textbf{\bibinfo {volume} {123}},\ \bibinfo {eid} {074305} (\bibinfo {year}
  {2005}),\
  \url{http://scitation.aip.org/content/aip/journal/jcp/123/7/10.1063/1.200165%
0}%
  \bibAnnoteFile{NoStop}{schinke1}%
\bibitem{schinke2}%
  \BibitemOpen
  \bibfield{author}{%
  \bibinfo {author} {\bibfnamefont{S.~Y.}\ \bibnamefont{Grebenshchikov}},
  \bibinfo {author} {\bibfnamefont{Z.-W.}\ \bibnamefont{Qu}}, \bibinfo {author}
  {\bibfnamefont{H.}~\bibnamefont{Zhu}},\ and\ \bibinfo {author}
  {\bibfnamefont{R.}~\bibnamefont{Schinke}},\ }%
  \bibfield{journal}{%
  \Doi{10.1039/B701020F}{\bibinfo {journal} {Physical Chemistry Chemical
  Physics}}\ }%
  \textbf{\bibinfo {volume} {9}},\ \bibinfo {pages} {2044} (\bibinfo {year}
  {2007}),\ \url{http://dx.doi.org/10.1039/B701020F}%
  \bibAnnoteFile{NoStop}{schinke2}%
\end{thebibliography}%

\end{document}